\documentclass{ws-mplb}

\newcommand{\half}{\frac{1}{2}}
\begin{document}
\markboth{R. K. Nesbet}{Beyond Density-Functional Theory:
The Domestication of Nonlocal Potentials}
\catchline{}{}{}{}{}
\title{BEYOND DENSITY FUNCTIONAL THEORY:\\
THE DOMESTICATION OF NONLOCAL POTENTIALS}
\author{\footnotesize ROBERT K. NESBET}
\address{IBM Almaden Research Center\\
650 Harry Road, San Jose, CA 95120, USA\\
rkn@almaden.ibm.com}
\maketitle
\begin{history}
\received{(received date)}
\revised{(revised date)}
\end{history}
\begin{abstract}
Due to efficient scaling with electron number N, density functional
theory (DFT) is widely used for studies of large molecules and solids.
Restriction of an exact mean-field theory to local potential functions
has recently been questioned.
This review summarizes motivation for extending current DFT
to include nonlocal one-electron potentials, and proposes
methodology for implementation of the theory.  The theoretical model,
orbital functional theory (OFT), is shown to be exact in principle for
the general N-electron problem.  In practice it must depend on a
parametrized correlation energy functional.  Functionals are proposed 
suitable for short-range Coulomb-cusp correlation and for long-range
polarization response correlation.  A linearized variational cellular 
method (LVCM) is proposed as a common formalism for molecules and 
solids.  Implementation of nonlocal potentials is reduced to 
independent calculations for each inequivalent atomic cell.
\end{abstract}
\section{Introduction}
Hohenberg and Kohn\cite{HAK64} (HK) proved for nondegenerate ground 
states that total electronic density determines the external
potential acting on an interacting N-electron system.  The constrained
search procedure of Levy\cite{LEV79} extends this proof to general
variational forms of N-electron wave functions, 
easily extended to spin-dependent densities. Spin indices and 
summations are assumed here, but omitted from notation.  
HK theory defines a universal ground state density 
functional $F_s=E-V$, which reduces for noninteracting electrons to
the kinetic energy functional $T_s$ of Kohn and Sham\cite{KAS65} (KS).
Implementation of the implied density-functional theory (DFT) followed 
only after introduction of the KS orbital mean-field model.
In KS theory, the density
function is $\rho({\bf r})=\sum_in_i\rho_i({\bf r})$, a sum of orbital 
densities $\rho_i=\phi^*_i\phi_i$ for the model state,
with occupation numbers $n_i$.
The many review articles and monographs on
DFT are exemplified by Parr and Yang\cite{PAY89} and by
Dreizler and Gross\cite{DAG90}.

\section{Motivation for an exact orbital theory}
\subsection{Can DFT with local potentials be an exact theory?}
The relationship of DFT to many-body theory remains obscure, requiring 
quite different methodology for strongly-correlated systems.
The KS construction is not easily related to strong-correlation
theory, in which a Hubbard model is often introduced that has no
relationship to the electronic density.
There is no systematic way to improve model functionals.
Nonphysical self-interaction produces incorrect ground states
for magnetically ordered systems.  Nonbonding interactions in molecules
and optical potentials in electron scattering are not correctly
described.
\par Numerous problems or paradoxes in DFT are associated with the
assumption that exact mean-field theory requires only local one-electron
potential functions\cite{NES03c}.  It is important to note that
the mathematical issue here is not the existence of HK 
functionals as such, but rather the existence of the uniquely defined
density functional derivatives required to determine variational
Euler-Lagrange equations\cite{NES03}.  Quantitative tests show that an
optimized effective potential (OEP local exchange) does not reproduce 
ground-state energies and densities in the unrestricted Hartree-Fock
(UHF) variational model for atoms\cite{ALT78,CAN01}.
Because the UHF energy functional is uniquely 
defined, and the accurately computable UHF ground-state density
is both noninteracting and interacting v-representable\cite{DAG90},
this textbook example of constrained variational theory is a rigorous
consistency test for DFT\cite{NES03}.  The exchange-only limit of DFT
linear response theory\cite{PGG96} is inconsistent with the
time-dependent Hartree-Fock theory of Dirac\cite{DIR30}, due to failure 
of locality\cite{NES99}.  A postulated local kinetic energy would imply
an exact Thomas-Fermi theory (TFT), in conflict with KS DFT and
with the exclusion principle for more than two electrons\cite{NES98}.
\par In an orbital model, such as KS theory, the exclusion 
principle is imposed by independent variation and normalization of each
occupied orbital of a model state\cite{NES03}.  This logic produces an
extended TFT\cite{NES02} when applied to orbital densities.  This
orbital-density theory is operationally equivalent to orbital 
functional theory (OFT), and reduces to KS DFT in the local
density approximation (LDA)\cite{NES02a}.  There is no conflict with
rigorous theory\cite{EAE84}.  Confined to normalized total densities,
functional analysis implies neither Thomas-Fermi nor KS 
equations\cite{NES02a}.
\subsection{OFT with nonlocal potentials can be an exact theory }  
For nondegenerate states, the KS Ansatz, $\rho=\sum_in_i\rho_i$, 
expresses $\rho$ as a function of occupied orbitals of a Slater
determinant model state $\Phi$.  With this Ansatz, any well-defined 
density functional is also an orbital functional.  If $F_s[\rho]=E-V$
is parametrized as an explicit functional of $\rho$, it defines
an orbital functional $F[\{\phi_i\}]$ for all functions in the orbital
Hilbert space.  Schr\"odinger variational theory is expressed
in terms of orbital functionals with functional differentials
of the form\cite{NES03}
\begin{eqnarray*}
\delta F = \sum_in_i\int d^3{\bf r} 
 \{\delta\phi^*_i\frac{\delta F}{n_i\delta\phi^*_i}+cc \}. 
\end{eqnarray*}
This defines an orbital functional derivative of the general form
$\frac{\delta F}{n_i\delta\phi^*_i}={\cal F}\phi_i$, where ${\cal F}$
is a linear operator.  These functional derivatives determine
orbital Euler-Lagrange (OEL) equations
${\cal F}\phi_i=\{\epsilon_i-v({\bf r})\}\phi_i$.
This is an exact theory if the true N-electron $F=E-V$ can be expressed 
as an orbital functional\cite{NES01a}.  
\par
OFT is deduced from formally exact N-electron theory, and nonlocal 
exchange-correlation potentials are constructed.  Quite explicitly,
the exchange "potential" is the Fock exchange operator, and the
correlation "potential" is determined by orbital functional derivatives
of the linked-cluster expansion\cite{NES03}.  The implied OEL equations
are a realization of "exact" KS equations.  OFT implies a
systematic refinement of the LDA based on implementation of nonlocal
potentials.  The theory is not density-dependent, bypassing questions
of v-representability and of the convergence of density-gradient
expansions.  It is not restricted in principle to ground states, to
weak correlation, or to short-range interactions\cite{NES01a}. 
It systematizes many alternatives currently being
pursued in molecular DFT: the orbital-dependent exchange originally
proposed by KS\cite{KAS65}, and the 
OEP approximation\cite{NES03,ALT78}, sometimes characterized
as "exact KS"\cite{KAP03}.  OFT nonlocal potentials are derived
from functionals defined by antisymmetric wave functions.  In principle,
this eliminates self-interaction and the need for any {\it ad hoc}
correction\cite{PAZ81}.  In practice, computational methodology
developed for the latter may be very useful in the context of the
orbital-indexed potentials considered here.  The present analysis has
important practical implications for computational methodology relevant
to the electronic structure of large molecules and solids.  With a new
theoretical basis embedded in N-electron theory, extension of current
work on exact exchange in DFT to include electronic correlation 
requires timely development of efficient computational methods
appropriate to nonlocal potentials\cite{NES03a}.

\section{Orbital functional theory}
\subsection{An exact mean-field theory exists}
Specify any relationship $\Psi\to\Phi$ that determines a model state 
$\Phi$ for any N-electron state $\Psi$, where $\Phi$ is a Slater
determinant for any nondegenerate state.  This postulate implies the 
KS Ansatz in a variational theory of the model density.
Imposing normalization $(\Phi|\Psi)=(\Phi|\Phi)=1$, which causes
no formal problems for a finite system,
$(H-E)\Psi=0$ implies that $E=E_0+E_c$, where $E_0=(\Phi|H|\Phi)$, and
$E_c=E-E_0=(\Phi|H|\Psi-\Phi)$ defines the correlation energy.
Since $E_0$ is an explicit orbital functional, an exact
mean-field theory exists if $E_c$ can be derived from many-body theory
as an orbital functional.  The linked-cluster expansion shows that 
this is possible.  Exact OEL equations are defined formally by
orbital functional derivatives of the linked-cluster expansion.
An implicit closed form of the exact functional $E_c$ has been
derived\cite{NES00}.
\subsection{Kinetic energy and exact exchange}
Using ${\hat t}=-\half\nabla^2$ and $v=v({\bf r})$,
functional $E_0=T+U+V$ is defined such that
\begin{eqnarray*}
T=\sum_in_i(i|{\hat t}|i);\;V=\sum_in_i(i|v|i)
\end{eqnarray*}
for orthonormal orbital functions.
The Coulomb interaction $u=1/r_{12}$ defines orbital functional 
$U=E_h+E_x$, where
\begin{eqnarray*}
E_h= \half\sum_{i,j}n_in_j(ij|u|ij);\;
E_x=-\half\sum_{i,j}n_in_j(ij|u|ji).
\end{eqnarray*}
\subsection{Variational equations for the orbitals}
Orbital Euler-Lagrange (OEL) equations are derived using the orbital
functional derivatives
\begin{eqnarray}\label{OFDs}
\frac{\delta T}{n_i\delta\phi^*_i}={\hat t}\phi_i;\;
\frac{\delta U}{n_i\delta\phi^*_i}={\hat u}\phi_i;\;
\frac{\delta V}{n_i\delta\phi^*_i}=v({\bf r})\phi_i.
\end{eqnarray}
Here ${\hat u}=v_h({\bf r})+{\hat v}_x$, where $v_h$ is the 
Coulomb potential, and ${\hat v}_x$ is the Fock exchange operator.
An orbital functional $E_c[\{\phi_i\}]$ defines
$\frac{\delta E_c}{n_i\delta\phi^*_i}={\hat v}_c\phi_i$.
It will be assumed that some parametrized $E_c$ is defined such that
${\hat v}_c$ is Hermitian.  A Lagrange functional consistent with
independent orbital normalization is defined by subtracting
$\sum_in_i[(i|i)-1]\epsilon_i$ from the energy functional.
Defining $F=E-V$, the OEL equations are
\begin{eqnarray}\label{OELeqs}
{\cal F}\phi_i=\{\epsilon_i-v\}\phi_i, \;\;i=1,\cdots,N,
\end{eqnarray}
using the orbital functional derivative
$\frac{\delta F}{n_i\delta\phi^*_i}={\cal F}\phi_i=
 \{{\hat t}+{\hat u}+{\hat v}_c\}\phi_i$.
Here ${\hat t}$ and ${\hat u}$ are explicitly nonlocal (linear
operators), and there is no implication that ${\hat v}_x+{\hat v}_c$ 
is equivalent to a multiplicative local potential.

\section{OFT as an exact variational model}
\subsection{Correlation energy as an orbital functional}
Unsymmetrical normalization 
$(\Phi|\Psi-\Phi)=0$ defines an orthogonal projection,
${\cal P}\Psi=\Phi$ and  ${\cal Q}\Psi=\Psi-\Phi$, where
${\cal P}=\Phi\Phi^\dagger$ and ${\cal Q}=I-{\cal P}$.
The N-electron Schr\"odinger equation can be partitioned 
by projection.  In the limit $\eta\to0+$,
\begin{eqnarray*}
{\cal Q}\Psi=-[{\cal Q}(H-E_0-E_c-i\eta){\cal Q}]^{-1}H\Phi.
\end{eqnarray*}
Hence, for $\eta\to0+$,
\begin{eqnarray*}
E_c&=&(\Phi|H|\Psi-\Phi)=(\Phi|H|{\cal Q}\Psi)\\
&=&-(\Phi|H[{\cal Q}(H-E_0-E_c-i\eta){\cal Q}]^{-1}H|\Phi).
\end{eqnarray*}
Because model state $\Phi$ determines ${\cal Q}$, this exact
expression for $E_c$ is an implicit orbital functional\cite{NES00}.
\subsection{G\^ateaux functional derivatives}
Suppose that an orbital functional $F$ is defined such that
$\frac{\delta F}{n_i\delta\phi^*_i}={\hat v}_f\phi_i$
and ${\hat v}_f$ is Hermitian in the orbital Hilbert space.
With no additional assumptions, the functional differential
\begin{eqnarray*}
\delta F &=&\sum_in_i\int d^3{\bf r}
\{\delta\phi^*_i({\bf r}){\hat v}_f\phi_i({\bf r})+cc\}
=\sum_in_i\int d^3{\bf r}
\frac{\phi^*_i{\hat v}_f\phi_i}{\phi^*_i\phi_i}\delta\rho_i 
\end{eqnarray*}
determines a density functional derivative.  This derivative
is an orbital-indexed local potential function
\begin{eqnarray}\label{Fgat}
\frac{\delta F}{n_i\delta\rho_i}=v_{fi}({\bf r})=
 \frac{\phi^*_i{\hat v}_f\phi_i}{\phi^*_i\phi_i},
\end{eqnarray}
defined throughout the orbital Hilbert space\cite{NES03a}.  
\par
For unconstrained orbital variations about any stationary state, the
functional differential of $F=E-V$ is determined by Eqs.(\ref{OELeqs}),
the OEL equations.  In detail\cite{NES01}, 
\begin{eqnarray*}
\delta F =\sum_in_i\int d^3{\bf r}
\{\delta\phi^*_i({\bf r}){\cal F}\phi_i({\bf r})+cc\}
=\sum_in_i\int d^3{\bf r}
\{\epsilon_i-v({\bf r})\}\delta\rho_i({\bf r}) 
\end{eqnarray*}
determines the density functional derivative
\begin{eqnarray*}
\frac{\delta F}{n_i\delta\rho_i}=v_{fi}({\bf r})
 =\epsilon_i-v({\bf r}). 
\end{eqnarray*}
Hence for any stationary state,
$v_{fi}({\bf r})$ depends on the orbital subshell index
unless all orbital energies are equal\cite{NES01}.
Such dependence on a "direction" in the Banach space of densities 
defines a G\^ateaux functional derivative, the generalization to 
functional analysis of an analytic partial derivative\cite{BAB92}.
The G\^ateaux derivative expressed by an indexed local potential
$v_{fi}({\bf r})$ is in general singular at nodes of the orbital
function $\phi_i({\bf r})$.  A Fr\'echet derivative\cite{BAB92} 
(multiplicative local potential) can exist only if $v_{fi}$ is 
independent of its orbital index at all coordinate points, which 
requires elimination or exact cancellation of the orbital nodes.  This 
condition is imposed by the construction of $v_{xc}$ in the LDA, but 
cannot be assumed to be a general consequence of variational theory.   
\par
In order to generate Euler-Lagrange equations with a normalization
constraint, a variational functional of total density must be 
defined for arbitrary infinitesimal variations about normalized 
densities\cite{NES03}.  The HK functional $F_s$ is defined only for 
normalized ground states.  No Euler-Lagrange equation is implied unless
this definition can be extended to include infinitesimal neighborhoods 
of ground states.  If a total (Fr\'echet) functional derivative
were to exist, it would justify such an extension, implying an
exact Thomas-Fermi theory\cite{NES02a}.  If the functional differential 
\begin{eqnarray*} 
\delta F=\int d^3{\bf r} \frac{\delta F}{\delta\rho}\delta\rho
\end{eqnarray*}
is a unique functional of the total density variation
$\delta\rho$, it defines a Fre\'chet functional derivative
\begin{eqnarray}\label{Ffre}
\frac{\delta F}{\delta\rho}=v_f({\bf r})
\end{eqnarray}
as a multiplicative local potential function.  Comparison of 
Eqs.(\ref{Ffre}) and (\ref{Fgat}) shows that the total functional
differential is not unique unless the corresponding G\^ateaux 
derivatives are all equal.  For variations 
about a ground state, this requires all orbital energies to be
equal, violating the exclusion principle for any compact 
system with more than one electron of each spin.
Hence exact TFT is not implied for more than two electrons.
In contrast, the G\^ateaux derivatives are well-defined, and imply
modified Thomas-Fermi equations with orbital indices, consistent
with the exclusion principle, and with the general form
of KS equations\cite{NES02}.  
\par Eqs.(\ref{Fgat}) and (\ref{Ffre}) cannot be reconciled for the
HK functional $F_s$ unless all orbital energy eigenvalues 
are equal\cite{NES98,NES02a}.  This difference is a consequence of 
different normalization constraints in DFT or OFT and in TFT.  
That independent normalization of orbital densities is physically 
correct, specifically enforcing the exclusion principle in Hartree-Fock 
theory and KS DFT, is obvious for the example of an atom with
two noninteracting electrons of parallel spin.  The ground state is
$1s2s\;^3S$.  The density constraints are:
\begin{eqnarray*}
DFT: \int\rho_{1s}=1,\;\int\rho_{2s}&=&1\\
TFT: \int(\rho_{1s}+\rho_{2s})&=&2.
\end{eqnarray*}
This TFT constraint allows the nonphysical solution
\begin{eqnarray*}
TFT: \int\rho_{1s}=2,\;\int\rho_{2s}&=&0,
\end{eqnarray*}
which violates the exclusion principle.

\section{How to do it}
The technical problem of generalizing KS DFT to an exact theory 
reduces to methodology for indexed local potentials.  
For atoms or local atomic cells, this just adds an indexed
correlation potential
$v_{ci}({\bf r})=\frac{\phi^*_i{\hat v}_c\phi_i}{\phi^*_i\phi_i}$ 
to standard Hartree-Fock methodology for the Fock exchange 
potential\cite{CFF77}.
Off-diagonal Lagrange multipliers can be used to constrain orbital
orthogonality during self-consistent iterations.
For closed-shell systems, these Lagrange multipliers must vanish
on convergence.
\subsection{Energy-linearized variational cellular method}
Basis-set expansion becomes impractical for large molecules and
solids.  The alternative methodology of multiple scattering theory,
used in many variants for the electronic structure of condensed
matter\cite{GAB00}, can be formulated as a variational cellular
method (VCM)\cite{FAL78,NES03}.  
In full-potential MST, local basis functions are computed numerically in
each atomic cell of a space-filling Voronoi lattice.  These functions 
are truncated at local cell boundaries, and linear combinations are 
matched variationally across cell interfaces\cite{NES03}.
The logical structure and inherent efficiency of the VCM emulate
a tight-binding model.  It can be linearized in energy and adapted
to a general variational method for solids and molecules\cite{NES03b}.
Using numerical basis functions in local cells validates
semirelativistic calculations for heavy atoms, and validates 
frozen-core approximations.  These features are exploited in
condensed-matter methodology.
Truncating basis functions at cell boundaries removes intercell
overlap and multicellular Coulomb integrals from the methodology.
Local cells interact through boundary matching and through 
extended potential functions that act as an external field in each cell.
As in quantum field theory, the essential physics is that of
quasiparticles interacting through effective mean fields. 
\subsection{The local nonlocal (LNL) model}
Intracell calculations including an indexed correlation potential
are just like Hartree-Fock for atoms, not a practical problem.
Indexed local potentials are consistent with MST, which simply
uses a different orbital basis in each local cell.  The residual
practical problem is the treatment of long-range tails of the indexed
potentials.  The truncated basis removes all intercell overlap
contributions to these potentials. 
Electronic correlation arises from two distinct causes.  The
short-range effect of the Coulomb cusp is significant for intracellular
correlation, while the long-range effect of polarization is 
expected to be the principal intercellular correlation effect.
In the case of metals, long-range correlation is required for
dynamical exchange screening, needed to eliminate nonphysical
behavior of exact exchange at the Fermi surface.
The long-range asymptotic form of the indexed potentials for
exchange and correlation is electrostatic, modified by polarization
response.
The significant success of DFT and of model theories suggests
that, not far outside a source charge density, the asymptotic
potentials in general reduce to nonindexed collective fields. 
If this is valid, it would justify a "local nonlocal" (LNL)
model, in which true indexed potentials are used only within local
cells, while asymptotic forms of the extracellular fields are combined
as an effective mean field in each such cell.
As in the Ewald expansion, this implies significant cancellation
or screening of fields.
This may justify neglecting the multipole part of the effective
field in each cell during self-consistent iterations, including it
as a first-order perturbation after convergence. 
\subsection{Coulomb-cusp correlation}
Short-range correlation is dominated by the Coulomb pole
$u=1/r_{12}$, requiring a $1+\half r_{12}$ cusp in the wave function.  
The Colle-Salvetti (CS) Ansatz\cite{CAS75} imposes this cusp condition,
multiplying model state $\Phi$ by a symmetrical factor
$\Pi_{i<j}[1-\xi({\bf r}_i,{\bf r}_j)]$, where, in relative coordinates 
${\bf q}={\bf r}_i-{\bf r}_j$ and ${\bf r}=\half({\bf r}_i+{\bf r}_j)$,
$\xi({\bf r},{\bf q})=
 \exp(-\beta^2q^2)[1-\Gamma({\bf r})(1+\half q)]$\cite{CAS75}.
Correlation energy can be expressed exactly as a sum of
pair-correlation energies,
$E_c=(\Phi|H|\Psi-\Phi)=\sum_{i<j}n_in_j\epsilon^c_{ij}$.
This can be parametrized, in analogy to CS, by
$\epsilon^c_{ij}=-(ij|{\bar u}\xi_{ij}(q)|ij)$,
where ${\bar u}=u(1-{\cal P}_{12})$, ${\cal P}_{12}$ is the 2-electron
exchange operator, and 
$\xi_{ij}(q)=\exp(-\beta^2_{ij}q^2)[1-\gamma_{ij}(1+\half q)].$
The normalization condition $(\Phi|\Psi-\Phi)=0$ determines
\begin{eqnarray*}
\gamma_{ij}=\frac{(ij|\exp(-\beta^2_{ij}q^2)|ij)}
 {(ij|\exp(-\beta^2_{ij}q^2)(1+\half q)|ij)},
\end{eqnarray*}    
such that $(ij|\xi_{ij}|ij)=0$ for each pair.
The free parameter $\beta_{ij}$ can be chosen to minimize
the Bethe-Goldstone (independent electron pair model) energy for
each orbital pair.
The nonlocal correlation potential is defined by
${\hat v}_c\phi_i=\frac{\delta E_c}{n_i\delta\phi^*_i}=
 -\sum_jn_j(j|{\bar u}\xi_{ij}|j)\phi_i.$
Antisymmetry is built in.  Terms $j=i$ vanish, so there is
no self-interaction.
The indexed local correlation potential is
\begin{eqnarray*}
v_{ci}({\bf r})=-\phi^{-1}_i({\bf r})
 \sum_jn_j(j|{\bar u}\xi_{ij}|j)\phi_i({\bf r}). 
\end{eqnarray*}
\subsection{Multipole response correlation}
If the mapping $\Psi\to\Phi$ is determined by the Brueckner-Brenig
condition\cite{BAW56,BRE57}, $(\delta\Phi|\Psi)=0$, so that $\Phi$ is 
the reference state of maximum weight in $\Psi$, this removes all 
single virtual excitations $\Phi_i^a$ from $\Psi$.  
Then the leading terms of ${\cal Q}\Psi$ are
double virtual excitations $\Phi_{ij}^{ab}$ with coefficients
$c_{ij}^{ab}$.  Higher-order terms do not interact with $\Phi$
through the Coulomb interaction.  Thus $E_c=(\Phi|H|{\cal Q}\Psi)$
is a sum of pair correlation energies\cite{NES03},
$\epsilon^c_{ij}=
\sum_{a<b}(1-n_a)(1-n_b)(ij|{\bar u}|ab)c_{ij}^{ab}$.
The expression $(a|{\hat v}_c|i)=(a|\delta E_c/n_i\delta\phi^*_i)$
can be evaluated, treating the coefficients $c_{ij}^{ab}$ as constants
in the current cycle of an iterative loop\cite{NES00}.
These matrix elements
\begin{eqnarray*}
(a|{\hat v}_c|i)&=&
 \sum_jn_j\sum_{c<b}(1-n_c)(1-n_b)(aj|{\bar u}|cb)(cb|{\bar c}|ij)
\\
&-&\sum_{k<j}n_k n_j\sum_b(1-n_b)(kj|{\bar u}|ib)(ab|{\bar c}|kj),
\end{eqnarray*}
agree with the leading self-energy diagrams in many-body
perturbation theory.  This determines the kernel of the linear
operator ${\hat v}_c$.
\par As an illustrative example, this formula has been simplified and
applied to the leading long-range term in an iterative expansion of
the multipole polarization potential implied by bound-free correlation
between a scattered electron and a polarizable target\cite{NES00}.
This analysis introduces a polarization pseudostate $\phi_{p_j}$,
computed as the normalized first-order response of an occupied 
orbital $\phi_j$ of the target system to an external multipole field. 
For multipole index $\lambda>0$, transition matrix elements
$(p_j |u|j)$ vary as $r^{\lambda}$ for small r and as $1/r^{\lambda+1}$
for $r\gg r_0$.  The implied multipole polarization potential is
quadratic in these transition elements and varies as
$r^{2\lambda}$ and $1/r^{2\lambda+2}$, respectively, in these limits.
Hence, unlike earlier models from purely asymptotic theory, this
potential vanishes as $r\to0$.
When applied for $\lambda=1$ to variational calculations of 
low-energy e-He scattering, 
this polarization potential produced
scattering phase shifts in good agreement with earlier variational
calculations that incorporated dipole polarization response.
The indicated general methodology is to compute first-order
multipole pseudostates within each atomic cell, and then use them
to construct the multipole polarization potentials that contribute
to the mean external field in all other cells.

\section{Conclusions}
This brief review is intended to motivate and to outline methodology 
that should be implemented in order to get beyond current practical 
limitations of density-functional theory.
It is a call to the condensed-matter community
to move beyond
DFT by extending current methodology for nonlocal potentials.
It is a call to the theoretical chemistry community,
concerned with
the electronic structure of large molecules, to set aside CI
methods and restricted DFT in favor of an efficient, practicable,
and ultimately exact {\it ab initio} mean-field theory.


\end{document}